\documentclass[aps,pre,twocolumn
,superscriptaddress,nofootinbib]{revtex4-1}

\usepackage{graphicx}
\usepackage{amsthm}
\usepackage[dvipsnames]{xcolor}
\usepackage{bm,amssymb,amsmath,mathtools,mathrsfs,url}
\allowdisplaybreaks

\begin{document}


\title{Bead-rod-spring models in random flows}



\author{Emmanuel Lance Christopher VI Medillo Plan}
\affiliation{Laboratoire Jean Alexandre Dieudonn\'e, Universit\'e Nice Sophia Antipolis, CNRS, 06108 Nice, France}

\author{Aamir Ali}
\affiliation{Laboratoire Jean Alexandre Dieudonn\'e, Universit\'e Nice Sophia Antipolis, CNRS, 06108 Nice, France}
\affiliation{Department of Mathematics, COMSATS Institute of Information Technology, Attock 43600, Pakistan}

\author{Dario Vincenzi}
\affiliation{Laboratoire Jean Alexandre Dieudonn\'e, Universit\'e Nice Sophia Antipolis, CNRS, 06108 Nice, France}



\date{\today}

\begin{abstract}
Bead-rod-spring models are the foundation of the kinetic theory of polymer solutions. We derive the diffusion equation for the probability density function of the configuration of a general bead-rod-spring model in short-correlated Gaussian random flows. Under isotropic conditions, we solve this equation analytically for the elastic rhombus model introduced by Curtiss, Bird, and Hassager [Adv. Chem. Phys. {\bf 35}, 31 (1976)].
\end{abstract}


\maketitle


The study of polymer solutions generally requires a coarse-grained description of a polymer molecule. 
A successful and well-established approach consists in using bead-rod-spring models, where a polymer is described as a sequence of beads connected by rigid or elastic links~\cite{BHAC77,DE86,O96,CCR12}.
By selecting the number of beads and the nature of the links, it is possible to build flexible, semi-flexible, or rigid molecules with various internal structures.
Bead-rod-spring models play a central role across several fields, including rheology, non-Newtonian fluid mechanics, chemical physics, soft matter~\cite{BHAC77,L98}. 
Analytical solutions of these models represent an essential step towards the derivation of constitutive equations and hence the prediction of the 
non-Newtonian properties of polymeric fluids~\cite{BHAC77,DE86,O96,L05}.
However, in spite of their conceptual simplicity, 
the internal dynamics of bead-rod-spring models may be exceedingly complex. 
For this reason bead-rod-spring  models have been solved analytically only in simplified settings. 
In the case of laminar flows, analytical results are restricted to linear velocity fields
with elementary time dependence, namely steady, start-up, or oscillatory extensional and shear flows~\cite{BHAC77,DE86,O96}.
In the case of randomly fluctuating flows, exact solutions are only available for dumbbells, which are simply composed of two beads and a single elastic or rigid 
link~\cite{SK92,BFL00,C00,T03,CMV05,CKLT05,MAV05,CPV06,VJBC07,T07,AV16}. 
Nevertheless, in recent years there has been a growing
interest in the Lagrangian dynamics of complex-shaped
objects in turbulent flows, such as elastic polymers \cite{WG10,LS14}, 
triaxial ellipsoids \cite{CM13},
flexible fibers \cite{BVL14}, 
crosses and jacks \cite{MPKNV14}, 
isotropic helicoids and four-bead particles \cite{GB15},
chiral dipoles \cite{KTTV16}.
It is therefore important to advance the analytical 
tools needed for the study of this problem.
Here we first derive the diffusion equation for 
the probability distribution of the configuration 
of a general bead-rod-spring polymer in a Gaussian 
random flow with short correlation time. 
Short-correlated stochastic fields have been 
widely employed in the theoretical study of turbulent 
flows and have yielded fundamental results 
on passive-scalar mixing
and the turbulent dynamo~\cite{FGV01}; moreover, they have been used to
predict the coil--stretch transition of polymers in turbulent flows~\cite{BFL00} and the associated critical slowing down~\cite{CPV06}.
By using the aforementioned diffusion equation,
we then analytically calculate the stationary configuration of the elastic rhombus model in an isotropic random flow.
The rhombus model was introduced by Curtiss, Bird, and Hassager~\cite{CBH76} and is a prototype of a finitely 
extensible multibead polymer that includes both elastic and rigid internal links.

We start by
briefly recalling the diffusion equation for the probability density function (PDF) of the configuration of a bead-rod-spring polymer following Refs.~\cite{BHAC77,O96}. 
Consider $N$ spherical beads with mass $m_\mu$, $\mu=1,\dots,N$. Let $\bm x_\mu$ denote the position vector of the $\mu$th bead with respect to a space-fixed coordinate system.
The position of the center of mass of the polymer is $\bm x_c=\sum_\mu m_\mu \bm x_\mu/
\sum_\mu m_\mu$, and the position vector of the $\mu$th bead referred to the center of mass is $\bm r_\mu=\bm x_\mu-\bm x_c$.
The polymer is immersed in a Newtonian fluid, whose motion is described by the incompressible velocity field $\bm u(\bm x,t)$.
The velocity gradient, $\bm\kappa=\nabla\bm u$, $\kappa^{ij}=\partial u^i/\partial x^j$,
is assumed to be uniform over the length of the polymer. 
If the flow is turbulent, this assumption means that the size of the polymer is much smaller than the viscous-dissipation scale.
The force of the flow on bead $\mu$ is given by Stokes's law with drag coefficient $\zeta_\mu$;
the inertia of the beads is disregarded. 
Furthermore, the concentration of the solution is sufficiently small for polymer--polymer hydrodynamic interactions to be negligible, and the flow is unperturbed by the presence of polymers.

Assume that the beads are subject to $D'$ rigidity constraints. Then, the number of degrees of freedom of the polymer in the frame of reference of the center of mass is $D=d(N-1)-D'$, where $d$ is the dimension of the flow. 
It is therefore convenient to specify the configuration of the polymer in terms of $D$ coordinates
$\bm q=(q^1,\dots,q^D)$, which describe its degrees of freedom.
The statistics of the coordinates $\bm q$ is given by the PDF $\psi(\bm q;t)$.

We define the tensors $\bm\zeta_{\mu\nu}$ implicitly from the equation:
$
\sum_\nu\bm\zeta_{\mu\nu}\cdot(\zeta_\nu^{-1}\delta_{\nu\mu'}\bm{\mathsf{I}}+
\bm{\mathsf{\Omega}}_{\nu\mu'})=\delta_{\mu\mu'}\bm{\mathsf{I}},
$
where $\delta_{\mu\mu'}$ is the Kronecker delta,
$\bm{\mathsf{I}}$ is the identity matrix, and the tensors
$\bm{\mathsf{\Omega}}_{\nu\mu}$ describe the 
hydrodynamic interactions between the $\mu$th and the $\nu$th bead
(in the simplest approximation $\bm{\mathsf{\Omega}}_{\mu\nu}$ is the Oseen tensor).
Define also the tensors
$\bm{\mathsf{Z}}=\sum_{\mu\nu}\bm\zeta_{\mu\nu}$, $\bm{\mathsf{\Lambda}}_\nu=
\bm{\mathsf{Z}}^{-1}\cdot\sum_\mu\bm \zeta_{\mu\nu}$, and
$\widetilde{\bm \zeta}_{\mu\nu}=
\bm\zeta_{\mu\nu}-\bm{\mathsf{\Lambda}}_\mu^{\mathrm{T}}\cdot
\bm{\mathsf{Z}}\cdot\bm{\mathsf{\Lambda}}_\nu$.
Finally, $\bm{f}_{\mu\nu}$ is the
force exerted by the $\nu$th bead over 
the $\mu$th one through the springs and $\bm{\mathfrak{f}}_\mu$
is the external force on bead $\mu$ (which is assumed to be independent
of $\bm x_c$).
Then, $\psi(\bm q;t)$ 
satisfies the diffusion equation~\cite{BHAC77,O96}
(summation over repeated indices is understood throughout):
\begin{multline}
\dfrac{\partial\psi}{\partial t}=
-\dfrac{\partial}{\partial q^i}\bigg\{
\widetilde{\mathsf{G}}^{ij}
\bigg[\big(\mathsf{M}^{jkl}\kappa^{kl}(t)
+F^j+\mathfrak{F}^j\big)\psi\\
-KT\sqrt{h}\dfrac{\partial}{\partial q^j}\bigg(\dfrac{\psi}{\sqrt{h}}\bigg)\bigg]
\bigg\},
\label{eq:diffusion}
\end{multline}
where $K$ is the Boltzmann constant, $T$ is temperature,
\begin{equation}
\mathsf{M}^{jkl}=r_\nu^l\dfrac{\partial r^m_\mu}{\partial q^j}\,
\widetilde{\zeta}_{\mu\nu}^{mk},\, 
F^j=\sum_\nu f^k_{\mu\nu}\dfrac{\partial r^k_\mu}{\partial q^j},\,
\mathfrak{F}^j=\mathfrak{f}^k_{\mu}\dfrac{\partial r^k_\mu}{\partial q^j},
\end{equation}
$h=\operatorname{det}(\bm{\mathsf{H}})$,
and $\widetilde{\bm{\mathsf{G}}}=\widetilde{\bm{\mathsf{H}}}^{-1}$ with:
\begin{equation}
\label{eq:H}
\mathsf{H}^{ij}=m_\mu\frac{\partial r^k_\mu}{\partial q^i}\frac{\partial 
r^k_\mu}{\partial q^j},\quad \widetilde{\mathsf{H}}^{ij}= \widetilde{\zeta}_{\mu\nu}^{kl} \,\frac{\partial r^k_\mu}{\partial q^i}\frac{\partial r^l_\nu}{\partial q^j}.
\end{equation}
The stationary solution of Eq.~\eqref{eq:diffusion}
can be calculated exactly when hydrodynamic bead--bead interactions are
negligible ($\bm{\mathsf{\Omega}}_{\mu\nu}=0$ for all $\mu,\nu=1,\dots,N$) 
or equilibrium averaged 
($\bm{\mathsf{\Omega}}_{\mu\nu}$ is replaced with its average value at equilibrium) 
and
the velocity gradient is time-independent and
symmetric ($\bm\kappa=\bm\kappa^\mathrm{T}$)~\cite{CBH76,BHAC77}.
For other flows, Bird \textit{et al.}~\cite{BHAC77} note that the analytical solution of 
Eq.~\eqref{eq:diffusion}
is in general a formidable problem.


Let us consider the case in which the velocity gradient fluctuates randomly in time, as in turbulent flows.
Assume that $\bm\kappa(t)$ is a delta-correlated-in-time
$(d\times d)$-dimensional Gaussian stochastic process with zero mean and correlation:
$\langle\kappa^{kl}(t)\kappa^{mn}(t')\rangle=                                                                             
\mathsf{K}^{klmn}\delta(t-t')$,
where the specific form of the tensor 
$\bm{\mathsf{K}}$
depends on the statistical symmetries of the flow.
Equation~\eqref{eq:diffusion} is now stochastic and describes the evolution of $\psi(\bm q;t)$ for a given realization of the flow; $\bm\kappa(t)$ is a multiplicative noise and is interpreted in the Stratonovich sense.
Under this assumption on the velocity gradient, the stochastic differential equation associated with 
Eq.~\eqref{eq:diffusion} is (see Ref.~\cite{G85}):
\begin{equation}
\label{eq:SDE}
d{q}^i=A^i\,dt+\mathsf{B}^{ij}\circ dW^j(t)+\mathsf{C}^{ikl}\circ d\Gamma^{kl}(t),
\end{equation}
where $\bm W(t)$ is $D$-dimensional Brownian motion and $\bm \Gamma(t)$ is such that $\bm\kappa(t)=d\bm\Gamma(t)/dt$, i.e. $\bm\Gamma(t)$ is a Gaussian process with $\langle\Gamma^{kl}(t)\rangle=0$ and $\langle\Gamma^{kl}(t)\Gamma^{mn}(t')\rangle=\mathsf{K}^{klmn}\min(t,t')$ \cite{FN1} . The symbol $\circ$ indicates that the stochastic differential equation is interpreted in the Stratonovich sense. The coefficients are
\begin{align}
&A^i=\widetilde{\mathsf{G}}^{ij}(F^j+\mathfrak{F}^j)
+KT\dfrac{\beta^{ia}}{\sqrt{h}}\dfrac{\partial}{\partial q^j}\big(\sqrt{h}\,\beta^{ja}\big),\\
&\mathsf{B}^{ij}=\sqrt{2KT}\beta^{ij},\quad\mathsf{C}^{ikl}=\widetilde{\mathsf{G}}^{ij}\mathsf{M}^{jkl}.
\end{align}
In the above equations, $\bm\beta$ is such that $\bm\beta\bm\beta^{\mathrm{T}}=
\widetilde{\bm{\mathsf{G}}}$ (it is assumed that $\widetilde{\bm{\mathsf{G}}}$ is positive definite---this condition is easily verified when the masses and the drag coefficients are the same for all beads). 
The It\^o form of Eq.~\eqref{eq:SDE} is (appendix \ref{app:derivation}):
\begin{multline}
\label{eq:Ito}
d{q}^i=\bigg(A^i+\frac{1}{2}\mathsf{B}^{ja}\frac{\partial \mathsf{B}^{ia}}{\partial q^j}
+\frac{1}{2}\mathsf{K}^{klmn}\mathsf{C}^{jmn}\dfrac{\partial\mathsf{C}^{ikl}}{\partial q^j}\bigg)\,dt\\
+\mathsf{B}^{ij}\,dW^j(t)+\mathsf{C}^{ikl}\,d\Gamma^{kl}(t).
\end{multline}
We denote by $p(\bm q;t)$ the PDF of the configuration of the polymer with respect to the realizations both of the velocity gradient and of thermal noise. The diffusion equation corresponding to Eq.~\eqref{eq:Ito} is (appendix \ref{app:derivation}):
\begin{multline}
\dfrac{\partial p}{\partial t}=
-\dfrac{\partial}{\partial q^i}\bigg[\bigg(A^i+
\frac{1}{2}\mathsf{B}^{ja}\frac{\partial\mathsf{B}^{ia}}{\partial q^j}
+\frac{1}{2}\mathsf{K}^{klmn}\mathsf{C}^{jmn}
\dfrac{\partial\mathsf{C}^{ikl}}{\partial q^j}\bigg)p\bigg]\\
+KT\dfrac{\partial^2}{\partial q^i\partial q^j}(\widetilde{\mathsf{G}}^{ij}p)
+\dfrac{\mathsf{K}^{klmn}}{2}\dfrac{\partial^2}{\partial q^i\partial q^j}
(\mathsf{C}^{ikl}\mathsf{C}^{jmn}p).
\label{eq:FPE}
\end{multline}
Rearranging the terms in Eq.~\eqref{eq:FPE}, we obtain:
\begin{multline}
\dfrac{\partial p}{\partial t}=
\dfrac{\partial}{\partial q^i}\Bigg\{\dfrac{1}{2}\mathsf{K}^{klmn}
\widetilde{\mathsf{G}}^{ia}\mathsf{M}^{akl}
\dfrac{\partial}{\partial q^j}\Big(
\widetilde{\mathsf{G}}^{jb}\mathsf{M}^{bmn} p\Big)\\
+\widetilde{\mathsf{G}}^{ij}
\Bigg[-(F^j+\mathfrak{F}^j)p
+KT\sqrt{h}
\dfrac{\partial}{\partial q^j}\Bigg(\dfrac{p}{\sqrt{h}}\Bigg)\Bigg]\Bigg\}.
\label{eq:random}
\end{multline}
Equation \eqref{eq:random} determines the evolution of the PDF of the configuration of a general bead-rod-spring model in a
short-correlated Gaussian flow.
Note that Eq.~\eqref{eq:random} can also be derived by applying Gaussian integration by parts~\cite{F95} to Eq.~\eqref{eq:diffusion}.

\begin{figure}[t]
\centering
\includegraphics[trim={1.9cm 1.8cm 1.7cm 1.2cm},clip,width=0.3\textwidth]{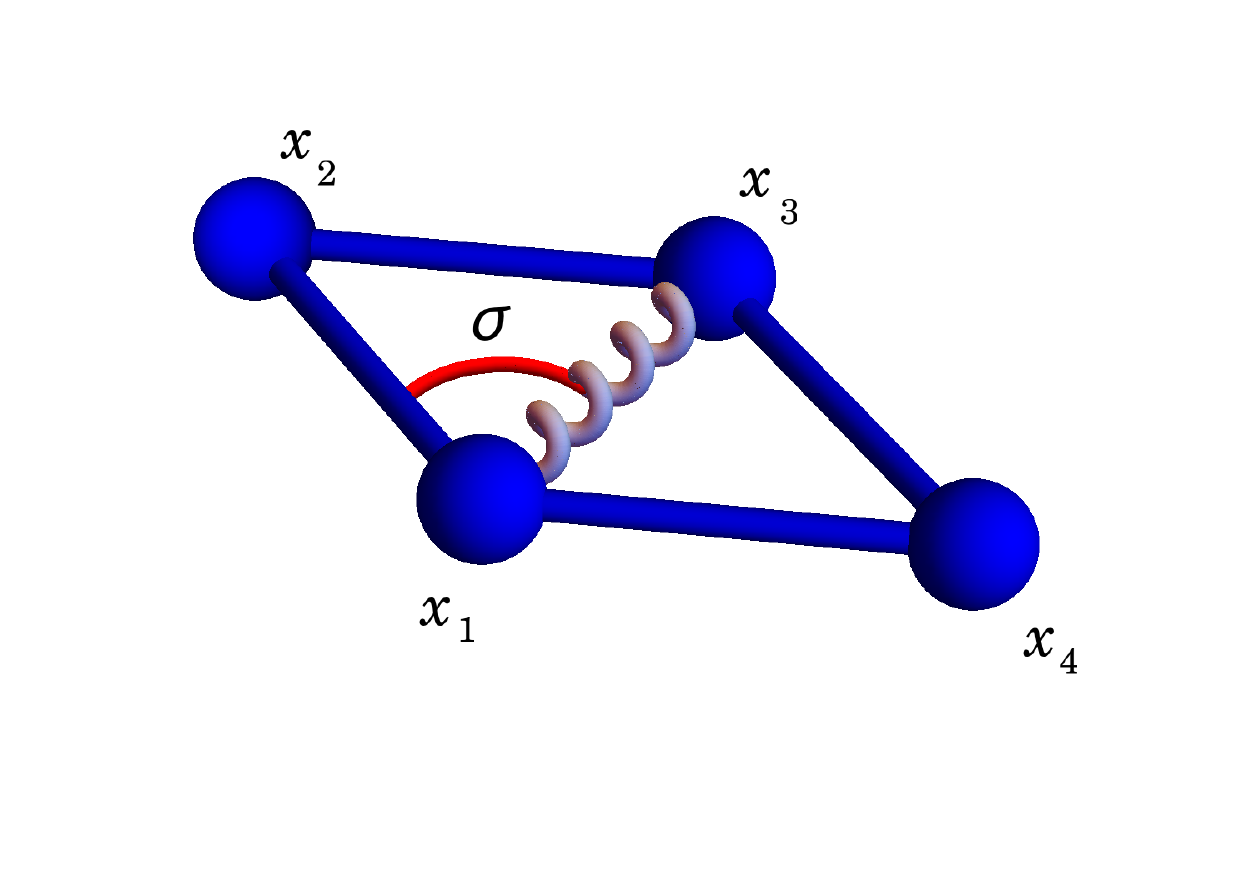}
\caption{The elastic plane rhombus.}
\label{fig:rhombus}
\end{figure}

We now consider a bead-rod-spring model for which the stationary solution of Eq.~\eqref{eq:random} can be calculated exactly.
The elastic plane rhombus model was introduced by Curtiss, Bird, and Hassager \cite{CBH76}. It describes a finitely extensible polymer and consists of four coplanar identical beads connected by four rods and of an elastic spring between two opposing beads (Fig.~\ref{fig:rhombus}). The angle $0\leq \sigma \leq \pi/2$ between the spring and one of the rods describes the deformation of the rhombus. The spring is at rest if the rods are perpendicular to each other. If $\sigma\neq \pi/4$, the spring stretches or compresses the rhombus back to its equilibrium; the force that it exerts on the beads is given by the harmonic potential $\phi(\sigma)=A(\sqrt{2}\cos\sigma-1)^2/2$.
Let $m_\mu=m$, $\zeta_\mu=\zeta$ for all $\mu=1,\dots,4$ and $\ell$ be the length of the rods. In addition, hydrodynamical bead--bead interactions are disregarded ($\bm{\mathsf{\Omega}}_{\mu\nu}=0$), and no external forces act on the polymer ($\bm{\mathfrak{f_\mu}}=0$). Under these assumptions, we have $\bm\zeta_{\mu\nu}=\zeta\delta_{\mu\nu}\bm{\mathsf{I}},\bm{\mathfrak{F}}=0$, and
\begin{equation}
\mathsf{M}^{jkl}=\zeta r_\mu^l\dfrac{\partial r^k_\mu}{\partial q^j},
\,\,F^j=-\frac{\partial \phi}{\partial q^j},
\,\,\widetilde{\mathsf{H}}^{ij}= \zeta \,
\frac{\partial r^k_\mu}{\partial q^i}
\frac{\partial r^k_\mu}{\partial q^j}.
\end{equation}
The contribution 
of the rhombic particles to the viscoelastic properties of a solution, such as the normal stress coefficient and the viscosity, have been calculated exactly~\cite{CBH76}.
In particular, the addition of $n$ rhombuses increases the viscosity by $3n\zeta \ell^2$. 
This result also holds for other variants of the rhombus model, namely the
freely-jointed and the rigid models. 
The normal stress coefficient of the polymer solution, by contrast, varies according to the flexibility of the rhombus. A fully elastic rhombus model has been used to investigate the motion of deformable active particles \cite{KLM16}. The rhombus model could also be used to examine how deformability influences the alignment and orientation statistics of microscopic particles in turbulent flows \cite{PCTV12}.

\begin{figure}[t]
\centering
\includegraphics[width=0.4\textwidth]{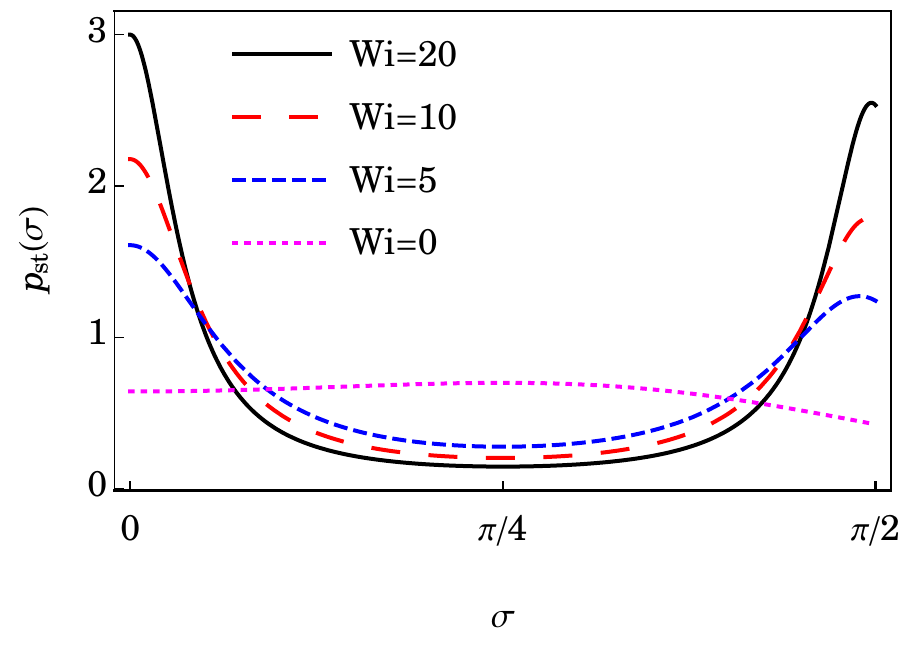}
\caption{Stationary PDF $p_\mathrm{st}(\sigma)$ for $d=2$ and $Z=1$, normalized such that $\int p_\mathrm{st}(\sigma)\, d\sigma=1$.}
\label{fig:pdf2D}
\end{figure}

If the random flow is incompressible and statistically invariant under rotations and reflections, the components of $\bm{\mathsf{K}}$ take the form:
${\mathsf{K}}^{ijkl}=2\lambda[(d+1)\delta^{ik}\delta^{jl}-\delta^{ij}
\delta^{kl}-\delta^{il}\delta^{jk}]/[d(d-1)]$, where $\lambda$ is the maximum Lyapunov exponent of the flow~\cite{FGV01}. Denoting $\bm{\mathsf{G}}=\zeta \bm{\widetilde{\mathsf{G}}},$ we write Eq.~\eqref{eq:FPE}
as a Fokker--Planck equation \cite{G85}: 
\begin{equation}
\dfrac{\partial p}{\partial t}=-\dfrac{\partial}{\partial q^i}(V^ip)+\dfrac{1}{2}\dfrac{\partial^2}{\partial q^i\partial q^j}(D^{ij}p),
\label{eq:FPErandomrhombus}
\end{equation}
where the drift and diffusion coefficients are
\begin{multline}
V^i=\dfrac{1}{2}\mathsf{K}^{klmn}\mathsf{G}^{jb}r_{\nu}^n\dfrac{\partial r_{\nu}^m}{\partial q^b}\dfrac{\partial}{\partial q^j}\Big(\mathsf{G}^{ia}r_\mu^l\dfrac{\partial r_{\mu}^k}{\partial q^a} \Big)\\
-\dfrac{\mathsf{G}^{ij}}{\zeta}\dfrac{\partial\phi}{\partial q^j}+\dfrac{KT}{\zeta\sqrt{h}}\dfrac{\partial}{\partial q^j}\Big(\mathsf{G}^{ij}\sqrt{h}\Big),
\label{eq:driftcoef}
\end{multline} 
and
\begin{equation}
D^{ij}=\mathsf{K}^{klmn}\mathsf{G}^{ia}r_{\mu}^l \dfrac{\partial r_{\mu}^k}{\partial q^a}
\mathsf{G}^{jb}r_{\nu}^n \dfrac{\partial r_{\nu}^m}{\partial q^b}
+\dfrac{2\mathsf{G}^{ij}KT}{\zeta}.
\label{eq:diffcoef}
\end{equation}
In the case of a two-dimensional flow ($d=2$), the rhombus may be described by $\bm q=(\theta,\sigma)$, where $\theta$ gives the orientation of $\bm x_3-\bm x_1$ with respect to the horizontal axis.
The vectors $\bm r_\mu$ can be expressed as:
\begin{eqnarray}
\bm r_1&=& \ell(-\cos\theta \cos\sigma,-\sin\theta \cos\sigma)=-\bm r_3,
\label{eq:r1}
\\
\bm r_2&=& \ell(-\sin\theta \sin\sigma,\cos\theta \sin\sigma)=-\bm r_4.
\label{eq:r2}
\end{eqnarray}
We also have $\mathsf{G}^{ij}=\delta_{ij}/2\ell^2$ and $h=4\ell^2$. After time is rescaled by the characteristic time scale of the restoring potential $\tau_\mathrm{p}=\zeta\ell^2/A$ and Eqs. \eqref{eq:r1}, \eqref{eq:r2} are replaced into Eqs.~\eqref{eq:driftcoef}, \eqref{eq:diffcoef}, the drift and diffusion coefficients in Eq.~\eqref{eq:FPErandomrhombus} become:
\begin{eqnarray}
V^\sigma &=&-(\sqrt{2}-2\cos\sigma)\sin\sigma/2,\\
D^{\theta\theta}&=& Z^{-1}+\mathrm{Wi}\,(5+\cos{4\sigma})/2,\\
D^{\sigma\sigma}&=& Z^{-1}+\mathrm{Wi}\,(1-\cos{4\sigma})/2,\\
V^{\theta}&=&D^{\sigma\theta}=D^{\theta\sigma}=0,
\end{eqnarray}
where $Z=A/KT$ is the stiffness parameter and the Weissenberg number $\mathrm{Wi}=\lambda\tau_\mathrm{p}$ measures the relative strength of the flow to that of the restoring potential $\phi(\sigma)$. 
The fact that all the coefficients are independent of $\theta$ reflects the isotropy of the flow and implies that the long-time PDF of the configuration is a function of $\sigma$ alone. Given reflecting boundary conditions at $\sigma=0$ and $\sigma=\pi/2$, the stationary PDF $p_{\mathrm{st}}(\sigma)$ can be analytically calculated to be \cite{G85}
\begin{equation}
 p_\mathrm{st}(\sigma)\propto\dfrac{1}{D^{\sigma\sigma}}\exp{\Big(2\int_0^\sigma\dfrac{V^{\sigma}(\omega)}{D^{\sigma\sigma }(\omega)}\,d\omega\Big)}.
\label{eq:stationarysol}
\end{equation}
We can evaluate the integral in Eq.~\eqref{eq:stationarysol} by
using Eqs.~(3.3.23) and (4.6.22) of Ref.~\cite{AS64}, whence the explicit form of the PDF of $\sigma$:
\begin{multline}
p_\mathrm{st}(\sigma)\propto\dfrac{1}{D^{\sigma\sigma}}
\left\lvert\frac{C_\sigma^2-C_+^2}{C_\sigma^2+C_-^2}\right\rvert^{Z/4C_1}\exp\left\{ \dfrac{Z\sqrt{\mathrm{Wi}\,Z}}{2C_1}\right.\\
\left.\times\left[\dfrac{\arctan\hspace{-1mm}\textrm{h}\left(C_\sigma/C_+\right)}{C_+}+\dfrac{\arctan\left(C_\sigma/C_-\right)}{C_-}\right]\right\}
\end{multline}
with
$C_\sigma=\sqrt{2\mathrm{Wi}\,Z}\cos\sigma$,
$C_1=\sqrt{\mathrm{Wi}\, Z(\mathrm{Wi}\,Z+1)}$, and
$C_\pm=\sqrt{C_1 \pm \mathrm{Wi}\,Z}$.
Figure \ref{fig:pdf2D} illustrates the graph of $p_{\mathrm{st}}(\sigma)$ for $Z=1$ and different values of $\mathrm{Wi}$.
 The probability of the rhombus being fully stretched $(\sigma=0)$ or fully compressed $(\sigma=\pi/2)$ increases with $\mathrm{Wi}$, i.e. the rhombus exhibits an almost rod-like configuration when a very strong flow dominates the dynamics. However, the stretched configuration is favored compared to the compressed one, because the restoring force is weaker for $\sigma=0$ than for $\sigma=\pi/2$.

\begin{figure}[t]
\centering
\includegraphics[width=0.4\textwidth]{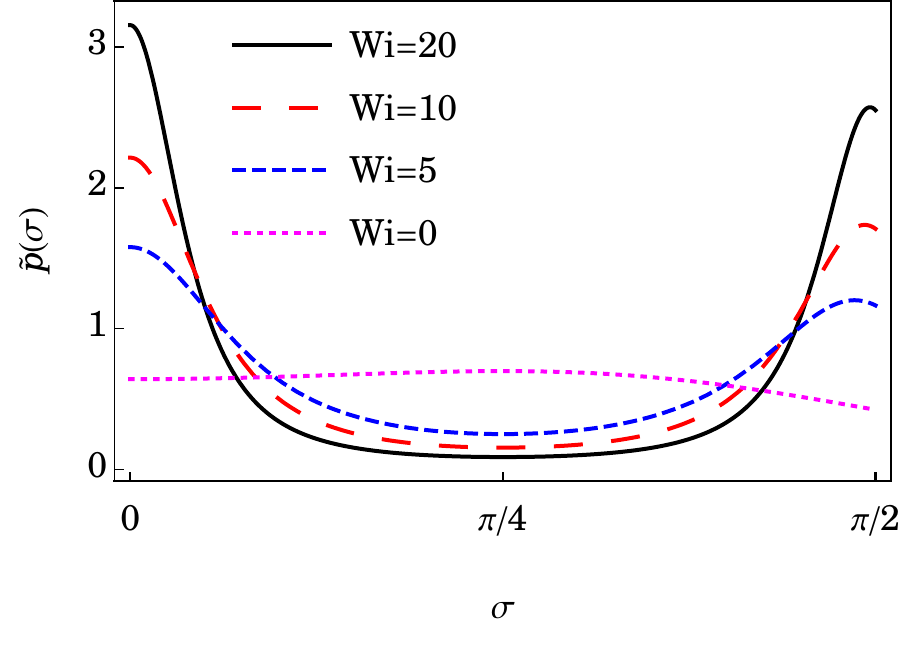}
\caption{The function $\tilde{p}(\sigma)$ for $d=3$ and $Z=1$, normalized such that $\int \tilde{p}(\sigma)\sin 2\sigma\, d\sigma=1$.}
\label{fig:pdf3D}
\end{figure}

In a three-dimensional flow, the number of degrees of freedom of the elastic planar rhombus is $D=4$. We consider $\bm{q}=(\alpha,\beta,\gamma,\sigma)$, where the first three coordinates are the Euler angles used to define the configuration of the rhombus in a fixed frame of reference. The details on the definition of the Euler angles, as well as the explicit expressions of $\mathsf G$ and $h$ can be found in Refs.~\citep{BHAC77,CBH76}. Note that our $\mathsf G$ differs from that of Ref.~\citep{CBH76} by a factor of $m$.
The PDF $p(\bm q)$ of the configuration again satisfies Eq.~\eqref{eq:FPErandomrhombus}, where $V^i$ and $D^{ij}$ are given in appendix \ref{app:coef}. In view of the isotropy of the flow, we assume that the stationary PDF of the configuration takes the form $p_\mathrm{st}(\bm q)=\tilde{p}(\sigma)\sin\beta\sin 2\sigma,$ where the factor $\sin\beta\sin 2\sigma$ is proportional to the Jacobian of the coordinate transformation from $\bm r_\mu$ to $\bm q.$ Substituting $p_\mathrm{st}(\bm q)$ into Eq.~\eqref{eq:FPErandomrhombus} results into a Fokker--Planck equation in the variable $\sigma$ alone:
\begin{equation}
0=\dfrac{\partial p_\mathrm{st}(\bm q)}{\partial s}=
-\dfrac{\partial}{\partial\sigma}(V^\sigma p_\mathrm{st})
+\dfrac{1}{2}\dfrac{\partial^2}{\partial\sigma^2}(D^{\sigma\sigma}p_\mathrm{st}),
\label{eq:FPErandomrhombusrescale3D}
\end{equation} 
where $s=t/\tau_\mathrm{p}$, $V^\sigma=\cot{2\sigma}/Z-(\sqrt{2}-2\cos\sigma)\sin\sigma/2$ and 
$D^{\sigma\sigma}=Z^{-1}+\mathrm{Wi}\,(1-\cos{4\sigma})/4.$
The partial derivatives with respect to the Euler angles indeed cancel each other. Under reflecting boundary conditions at $\sigma=0$ and $\sigma=\pi/2$, the solution of Eq.~\eqref{eq:FPErandomrhombusrescale3D} takes the form given in Eq.~\eqref{eq:stationarysol} with modified expressions for $V^\sigma$ and $D^{\sigma\sigma}$ \cite{G85}. Invoking again Eqs.~(3.3.23) and (4.6.22) of Ref.~\cite{AS64}, we derive the exact form of the stationary PDF of $\sigma$ in three dimensions:
\begin{multline}
\tilde{p}(\sigma)\propto\dfrac{1}{D^{\sigma\sigma}\sqrt{\mathrm{Wi}\, Z \sin^2 2\sigma+2}}
\left\lvert\frac{C_\sigma^2-C_+^2}{C_\sigma^2+C_-^2}\right\rvert^{Z/4C_1} \\
\times \exp\left\{\dfrac{Z\sqrt{\mathrm{Wi}\,Z/2}}{2C_1}\left[\dfrac{\arctan\hspace{-1mm}\textrm{h}\left(C_\sigma/C_+\right)}{C_+}  
\right. \right. \\
\left.\left.+\dfrac{\arctan \left(C_\sigma/C_-\right)}{C_-} \right] \right\} \end{multline}
with
$C_\sigma=\sqrt{\mathrm{Wi}\,Z}\cos\sigma$,
$C_1=\sqrt{\mathrm{Wi}\,Z(\mathrm{Wi}\,Z+2)}/2$, and
$C_\pm=\sqrt{C_1\pm\mathrm{Wi}\,Z/2}$.
The graph of $\tilde{p}(\sigma)$ is shown in Fig.~\ref{fig:pdf3D}. 
The statistics of the internal angle $\sigma$ is qualitatively the same for $d=2$ and $d=3$ (the curves in Figs.~2 and 3 indeed differ only slightly for the same Wi). This behavior is attributed to the fact that the rhombus can only deform in the plane to which it belongs and cannot undergo three-dimensional deformations, which makes its dynamics weakly sensitive to the dimension of the flow.

Bead-rod-spring models are the foundation of the kinetic theory of polymer solutions.
There is a broad literature on the analytic solutions of these models for laminar flows. In the case of fluctuating flows, such as turbulent flows, analytical results are restricted to dumbbells.
We have given a general description of bead-rod-spring models in short-correlated random flows. In addition, we have exactly solved the elastic rhombus model under isotropic conditions.
To the best of our knowledge, this is the first instance of an exact solution of a multibead model that include both elastic and rigid links and is transported by a randomly fluctuating flow.
We hope that the tools developed here will stimulate new studies on the dynamics of complex-shaped particles in turbulent flows.
The stretching dynamics of bead-spring chains has also received attention in the context of nonequilibrium statistical mechanics \cite{D05}; the probability of work and entropy production have been studied for laminar gradient flows \cite{TCCP07,SC11,LHS14,VTC15}. 
It would be interesting to generalize those studies to the case considered here, where the velocity gradient is a tensorial noise (see, e.g., Ref.~\cite{CG08}).

\begin{acknowledgments}
The authors are grateful to S.S. Ray for useful suggestions.
The work of AA and ELCMP was supported by EACEA through the Erasmus Mundus Mobility with Asia program. ELCMP and DV acknowledge the support of the EU COST Action MP 1305 
``Flowing Matter.''
\end{acknowledgments}

\appendix
\section{Derivation of Eqs.~\eqref{eq:Ito} and \eqref{eq:FPE}}
\label{app:derivation}

\noindent
The It\^o version of Eq.~\eqref{eq:SDE} can be derived formally by generalizing the standard transformation rules for stochastic differential equations (see, e.g., Refs.~\citep{G85,KP99}) to tensorial Brownian motion. Let us assume that the It\^o version is:
\begin{equation}
d{q}^i(t)=\mathfrak{A}^i\,dt+\mathfrak{B}^{ij} \,dW^j(t)+\mathfrak{C}^{ikl}\,d\Gamma^{kl}(t),
\label{eq:itotoderive}
\end{equation}
where the coefficients $\bm{\mathfrak{A,B,C}}$ are to be determined.
We first write Eq.~\eqref{eq:SDE} in integral form:
\begin{multline}
q^i(t)=q^i(t_0)+\int_{t_0}^tA^i\,dt'+\int_{t_0}^t\mathsf{B}^{ia}\circ\,dW^a(t')
\\+\int_{t_0}^t\mathsf{C}^{ikl}\circ\,d\Gamma^{kl}(t').
\label{eq:itointegralform}
\end{multline}
Consider a partition of $[t_0,t]$ into $N$ subintervals $[t_{\tau-1},t_\tau], \tau=1,\ldots,N$. The Stratonovich integrals in Eq.~\eqref{eq:itointegralform} are, by definition:
\begin{eqnarray}
\int_{t_0}^t\mathsf{B}^{ia}\circ\,dW^a(t')
&=& \lim_{N\rightarrow\infty}\sum_{\tau=1}^N \mathsf{B}^{ia}
\left(\bm q^{\mathrm{S}}_\tau,t_{\tau-1}\right)\Delta W^a_\tau, \qquad
\label{eq:integB}\\
\int_{t_0}^t\mathsf{C}^{ikl}\circ\,d\Gamma^{kl}(t')
&=& \lim_{N\rightarrow\infty}\sum_{\tau=1}^N \mathsf{C}^{ikl}
\left(\bm q^{\mathrm{S}}_\tau,t_{\tau-1}\right)\Delta\Gamma_\tau^{kl},
\label{eq:integC}
\end{eqnarray}
where
the limits are understood in the mean-square, 
$\bm q^{\mathrm{S}}_\tau=(\bm q_\tau+\bm q_{\tau-1})/2$, $\bm q_\tau=\bm q(t_\tau)$, $\Delta W_\tau^a=W^a(t_\tau)-W^a(t_{\tau-1})$, and $\Delta\Gamma_\tau^{kl}=\Gamma^{kl}(t_\tau)-\Gamma^{kl}(t_{\tau-1})$. We also introduce the notations $\Delta t_\tau=t_\tau-t_{\tau-1}$ and 
$\mathsf{X}_\tau^{I}=\mathsf{X}^I(\bm q_\tau,t_\tau)$ for any tensor $\mathsf{X}$ and set of indices $I$.
Define $\Delta\bm q_{\tau-1}=\bm q_\tau-\bm q_{\tau-1}$, whence
$\bm q^{\mathrm{S}}_\tau
=\bm q_{\tau-1}+\Delta\bm q_{\tau-1}/2.$
Expanding the coefficients in \eqref{eq:integB} and \eqref{eq:integC} at $\bm q_{\tau-1}$ 
yields:
\begin{eqnarray}
\mathsf{B}^{ia}(\bm q^{\mathrm{S}}_\tau,t_{\tau-1})
&=&\mathsf{B}^{ia}_{\tau-1}
+ \frac{\Delta q^j_{\tau-1}}{2}\frac{\partial\mathsf{B}^{ia}_{\tau-1}}{\partial q^j} 
+\mathrm{h.o.t.},\quad \label{eq:B2}\\
\mathsf{C}^{ikl}(\bm q^{\mathrm{S}}_\tau,t_{\tau-1})
&=&\mathsf{C}^{ikl}_{\tau-1}
+ \frac{\Delta q^j_{\tau-1}}{2} \frac{\partial\mathsf{C}^{ikl}_{\tau-1}}{\partial q^j} 
+\mathrm{h.o.t.}, \label{eq:C2}
\end{eqnarray}
where h.o.t. refers to higher order terms.
We now write $\Delta q^j_{\tau-1}$ according to its It\^o discretization:
\begin{equation}
\Delta q^j_{\tau-1}=\mathfrak{A}^j_{\tau-1}\Delta t_\tau+\mathfrak{B}^{ja}_ {\tau-1}\Delta W_\tau^a+\mathfrak{C}^{jmn}_{\tau-1}\,\Delta\Gamma_\tau^{mn}+\mathrm{h.o.t.}\quad
\label{eq:itolemma}
\end{equation}
After substituting Eqs.~\eqref{eq:B2}, \eqref{eq:C2}, \eqref{eq:itolemma} into Eqs.~\eqref{eq:integB}, \eqref{eq:integC} and using the formal rules $\Delta\Gamma^{kl}\times \Delta\Gamma^{mn}=\mathsf{K}^{klmn}\Delta t$, $\Delta W^j\times \Delta W^a=\delta^{ja}\Delta t$ and $\Delta W^{j}\times \Delta\Gamma^{kl}=0$ (owing to the independence of the velocity gradient and thermal noise), we can write Eqs.~\eqref{eq:integB} and  \eqref{eq:integC} as:
\begin{eqnarray}
\int_{t_0}^t\mathsf{B}^{ia}\circ\,dW^a(t')
&=&\lim_{N\rightarrow\infty}\sum_{\tau=1}^N  (\mathsf{B}^{ia}_{\tau-1}\,\Delta W_\tau^a 
\nonumber\\&&
+\frac{\mathfrak{B}^{ja}_ {\tau-1}}{2}\frac{\partial\mathsf{B}^{ia}_{\tau-1}}{\partial q^j}\,\Delta t), \nonumber\\
\int_{t_0}^t\mathsf{C}^{ikl}\circ\,d\Gamma^{kl}(t')
&=&\lim_{N\rightarrow\infty}\sum_{\tau=1}^N  \bigg(\mathsf{C}^{ikl}_{\tau-1}\,\Delta\Gamma_\tau^{kl}\nonumber\\
&&+\frac{1}{2}\mathsf{K}^{klmn}\mathfrak{C}^{jmn}_{\tau-1}\frac{\partial\mathsf{C}^{ikl}_{\tau-1}}{\partial q^j}\,\Delta t\bigg).\nonumber
\end{eqnarray}
The first terms in the sums yield It\^o stochastic integrals, whereas the second terms give ordinary integrals, hence:
\begin{eqnarray}
\int_{t_0}^t\mathsf{B}^{ia}\circ\,dW^a(t')
&=&\int_{t_0}^t\mathsf{B}^{ia}\,dW^a(t')
+\int_{t_0}^t\frac{\mathfrak{B}^{ja}}{2}\frac{\partial\mathsf{B}^{ia}}{\partial q^j}\,dt',\nonumber\\
\int_{t_0}^t\mathsf{C}^{ikl}\circ\,d\Gamma^{kl}(t')
&=&\int_{t_0}^t\mathsf{C}^{ikl}\,d\Gamma^{kl}(t') \nonumber\\
&&\qquad\qquad
+\int_{t_0}^t\mathsf{K}^{klmn}\mathfrak{C}^{jmn}\frac{\partial\mathsf{C}^{ikl}}{\partial q^j}\,dt'.\nonumber
\end{eqnarray}
Therefore, for Eq.~\eqref{eq:itotoderive} to be equivalent to Eq.~\eqref{eq:SDE}, the coefficients must satisfy $\mathfrak{B}=\bm{\mathsf{B}}$, $\mathfrak{C}=\bm{\mathsf{C}},$ and
\begin{equation}
\mathfrak{A}^i=A^i
+\frac{1}{2}\mathsf{B}^{ja}\frac{\partial\mathsf{B}^{ia}}{\partial q^j}
+\frac{1}{2}\mathsf{K}^{klmn}\mathsf{C}^{jmn}\frac{\partial\mathsf{C}^{ikl}}{\partial q^j}.
\nonumber
\end{equation}
We now derive the diffusion equation~\eqref{eq:FPE}.
Let $f(\bm q)$ be a function of $\bm q$. We first generalize It\^o's lemma for a tensorial Brownian motion. By expansion, we have:
\begin{equation}
f(\bm q+\Delta\bm q)-f(\bm q)
=\Delta q^i\frac{\partial f}{\partial q^i}+\frac{\Delta q^i\Delta q^j}{2}\frac{\partial^2 f}{\partial q^i\partial q^j}+\mathrm{h.o.t.} \nonumber
\end{equation}
By substituting $\Delta q^i$ and $\Delta q^j$ from Eq.~\eqref{eq:itotoderive} and using the formal rules as above, we obtain:
\begin{multline}
\frac{df}{dt}=\bigg[\mathfrak{A}^i+\mathfrak{B}^{ia}\frac{dW^a(t)}{dt}+\mathfrak{C}^{ikl}\frac{d\Gamma^{kl}(t)}{dt}\bigg]\frac{\partial f}{\partial q^i}\\
+\frac{1}{2}\left(\mathfrak{B}^{ia}\mathfrak{B}^{ja}
+\mathsf{K}^{klmn}\mathfrak{C}^{ikl}\mathfrak{C}^{jmn}\right)
\frac{\partial^2 f}{\partial q^i\partial q^j}.
\nonumber
\end{multline}
By taking the average over the noises and by using the properties of the It\^o integral, we have:
\begin{multline}
\left\langle\frac{df}{dt}\right\rangle
=\frac{d}{dt}\langle f\rangle
=\int f\left(\frac{\partial p}{\partial t}\right)\,d\bm q, \\
=\int \bigg[\mathfrak{A}^i\frac{\partial f}{\partial q^i}+\frac{1}{2}\left(\mathfrak{B}^{ia}\mathfrak{B}^{ja}
+\mathsf{K}^{klmn}\mathfrak{C}^{ikl}\mathfrak{C}^{jmn}\right)\frac{\partial^2 f}{\partial q^i\partial q^j}\bigg]p \,d\bm q.
\nonumber
\end{multline}
Integrating the right-hand side by parts gives:
\begin{multline}
\int f \bigg(\frac{\partial p}{\partial t}\bigg)\,d\bm q
= \int f\bigg[-\frac{\partial}{\partial q^i}(\mathfrak{A^i}p)+\frac{1}{2}\frac{\partial^2}{\partial q^i \partial q^j}\bigg(\mathfrak{B}^{ia}\mathfrak{B}^{ja}p\\[0pt]
+\mathsf{K}^{klmn}\mathfrak{C}^{ikl}\mathfrak{C}^{jmn}p\bigg)\bigg]\,d\bm q.
\nonumber
\end{multline}
Since $f(\bm q)$ is arbitrary, the expressions inside the brackets should be equal. This gives Eq.~\eqref{eq:FPE} when $\bm{\mathfrak{A,B}}$ and $\bm{\mathfrak{C}}$ are chosen as above.
\medskip

\section{Coefficients of the diffusion equation}
\label{app:coef}

\noindent
In this appendix, we give the 
the drift and diffusion coefficients of Eq.~\eqref{eq:FPErandomrhombus} for the 
three-dimensional elastic rhombus model. 
After the time variable is rescaled by $\tau_\mathrm{p}$, the drift coefficients $V^i$ become:
\begin{eqnarray}
V^\alpha &=& 2\sin{2\gamma}\cot\beta\csc\beta\cot{2\sigma}\csc{2\sigma}/Z, \nonumber \\
V^\beta &=& \cot\beta\csc^2{2\sigma}\left[(1+\cos{2\gamma}\cos{2\sigma})/Z  \right. \nonumber \\
&&  \left. +\mathrm{Wi}(1-\cos{4\sigma})/3 \right],  \nonumber\\
V^\gamma &=& -(3+\cos{2\beta})\csc^2{\beta}\cot{2\sigma}\csc{2\sigma}\sin{2\gamma}/2Z, \nonumber\\
V^\sigma &=& \cot{2\sigma}/Z-(\sqrt{2}-2\cos\sigma)\sin\sigma/2. \nonumber
\end{eqnarray}
The diffusion coefficients $D^{ij}$ are:
\begin{eqnarray}
D^{\alpha\alpha}&=& 4\csc{2\beta} \, V^{\beta}, \nonumber\\
D^{\alpha\beta}&=& -\tan\beta \, V^{\alpha}, \nonumber\\
D^{\alpha\gamma}&=& -2\csc{\beta} \, V^{\beta}, \nonumber\\
D^{\beta\beta}&=& (\cos^2{\gamma}\sec^2{\sigma}+\sin^2{\gamma}\csc^2{\sigma})/Z+4\mathrm{Wi}/3 ,\nonumber\\
D^{\beta\gamma}&=& \sin\beta \, V^\alpha, \nonumber\\
D^{\gamma\gamma}&=& 48\csc^2{2\sigma}\left\{24\left[4 \left(\cos{2\gamma}\cos{2\sigma}\cot^ 2{\beta}+\csc^2{\beta} \right.\right.\right. \nonumber \\
&& \left.\left. \left.-\cos{4\sigma} \right)-3\right]/Z+\mathrm{Wi}\left[-3\cos{8\sigma}+12\cos{4\sigma} \right.\right. \nonumber \\
&& \left. \left. 
+32\csc^2{\beta}\left(1-\cos{4\sigma}\right)-9  \right]  \right\},\nonumber \\
D^{\sigma\sigma}&=& Z^{-1}+\mathrm{Wi}\,(1-\cos{4\sigma})/4. \nonumber
\end{eqnarray}
By symmetry, $D^{\beta\alpha}= D^{\alpha\beta},D^{\gamma\alpha}= D^{\alpha\gamma},D^{\gamma\beta}= D^{\beta\gamma}$. All the other coefficients are zero.


\begin{thebibliography}{10}
\bibitem{BHAC77}
R.B. Bird, C.F. Curtiss, R.C. Armstrong, and O. Hassager,
\textit{Dynamics of Polymeric Liquids} (John Wiley and Sons, Inc., 1977), Vol.~2.

\bibitem{DE86}
M. Doi and S.F. Edwards, \textit{The Theory of Polymer Dynamics} (Oxford University Press, 1988).

\bibitem{O96}
H.C. \"{O}ttinger, 
\textit{Stochastic Processes in Polymeric Fluids} (Springer, Berlin, Germany, 1996).

\bibitem{CCR12}
C. Cruz, F. Chinesta, and G. R{\'e}igner,
Arch. Comput. Methods Eng. \textbf{19}, 227 (2012).

\bibitem{L98}
R.G. Larson, \textit{The Structure and Rheology of Complex Fluids} (Oxford University Press, 1999).

\bibitem{L05}
R.G. Larson, 
J. Rheol. \textbf{49}, 1 (2005).

\bibitem{SK92}
E.S.G. Shaqfeh and D.L. Koch, 
J. Fluid Mech. \textbf{244}, 17 (1992).

\bibitem{C00}
M. Chertkov, 
Phys. Rev. Lett. \textbf{84}, 4761 (2000).

\bibitem{BFL00}
E. Balkovsky, A. Fouxon and V. Lebedev,
Phys. Rev. Lett. \textbf{84}, 4765 (2000).
	
\bibitem{T03}
J.-L. Thiffeault,
Phys. Lett. A \textbf{308}, 445 (2003).

\bibitem{CMV05}
A. Celani, S. Musacchio, and D. Vincenzi,
J. Stat. Phys. \textbf{118}, 531 (2005).

\bibitem{CKLT05}
M. Chertkov, I. Kolokolov, V. Lebedev, and K. Turitsyn,
J. Fluid Mech. \textbf{531}, 251 (2005).

\bibitem{MAV05}
M. Martins Afonso and D. Vincenzi,
J. Fluid Mech. \textbf{540}, 99 (2005).

\bibitem{CPV06}
A. Celani, A. Puliafito, and D. Vincenzi,
Phys. Rev. Lett. \textbf{97}, 118301 (2006).

\bibitem{VJBC07}
D. Vincenzi, S. Jin, E. Bodenschatz, and L.R. Collins,
Phys. Rev. Lett. \textbf{98}, 024503 (2007).

\bibitem{T07}
K.S. Turitsyn,
J. Exp. Theor. Phys. \textbf{105}, 655 (2007).

\bibitem{AV16}
A. Ahmad and D. Vincenzi,
Phys. Rev. E \textbf{93}, 052605 (2016).

\bibitem{WG10}
T. Watanabe and T. Gotoh,
Phys. Rev. E \textbf{81}, 066301 (2010).

\bibitem{LS14}
Y. Liu and V. Steinberg,
Macromol. Symp. \textbf{337}, 34 (2014).

\bibitem{CM13}
L. Chevillard  and C. Meneveau,
J. Fluid Mech. \textbf{737}, 571 (2013).

\bibitem{BVL14}
C. Brouzet, G. Verhille, and P. Le Gal,
Phys. Rev. Lett. \textbf{112}, 074501 (2014).

\bibitem{MPKNV14}
G.G. Marcus, S. Parsa, S. Kramel, R. Ni, and G.A. Voth,
New J. Phys. \textbf{16}, 102001 (2014).

\bibitem{GB15}
K. Gustavsson and L. Biferale,
Proceedings of the 15th European Turbulence Conference (25--28 August, Delft, 2015), N. 103.
\path{http://www.etc15.nl/proceedings/proceedings/documents/103.pdf}.

\bibitem{KTTV16}
S. Kramel, S. Tympel, F. Toschi, and G.A. Voth,
\path{http://arxiv.org/abs/1602.07413}.

%
%



\bibitem{FGV01}
G. Falkovich, K. Gaw\c{e}dzki, and M. Vergassola, 
Rev. Mod. Phys. \textbf{73}, 913 (2001).

\bibitem{CBH76}
C.F. Curtiss, R.B. Bird, and O. Hassager, in \textit{Advances in Chemical Physics} \textbf{35}, edited
by I. Prigogine and S.A. Rice (John Wiley \& Sons, NJ, USA, 1976), pp. 31-117.

\bibitem{FN1}
Equation~\eqref{eq:SDE} may also be written as:
$d{q}^i=A^i\,dt+\mathsf{Q}^{ij}\circ d{\eta}^j(t)$, 
where $\bm\eta$ is $D$-dimensional Brownian motion and $(\mathsf{QQ}^\top)^{ij}=\mathsf{B}^{ia}\mathsf{B}^{ja}
+\mathsf{K}^{klmn}\mathsf{C}^{ikl}\mathsf{C}^{jmn}$. The thermal and velocity fluctuations would thus be merged into a single noise and the prefactor $\mathsf{Q}^{ij}$ would contain the information on the correlations of the random medium.

\bibitem{G85}
C.W. Gardiner, \textit{Handbook of Stochastic Methods} (Springer, Berlin, Germany, 1985).

\bibitem{F95}
U. Frisch, \textit{Turbulence: The Legacy of A. N. Kolmogorov} (Cambridge  University  Press, 1995).

\bibitem{KLM16}
N. K\"uchler, H. L\"owen, and A. M. Menzel, 
Phys. Rev. E \textbf{93}, 022610 (2016).

\bibitem{PCTV12}
S. Parsa, E. Calzavarini, F. Toschi, and G.A. Voth,
Phys. Rev. Lett. \textbf{109}, 134501 (2012).

\bibitem{AS64}
M. Abramowitz and I.~A. Stegun, \textit{Handbook of Mathematical Functions} (National Bureau of Standards, U.S. Government Printing Office, 1964).

\bibitem{D05}
A. Dhar,
Phys. Rev. E {\bf 71}, 036126 (2005).

\bibitem{TCCP07}
K. Turitsyn, M. Chertkov, V.Y. Chernyak, and A. Puliafito,
Phys. Rev. Lett. {\bf 98}, 180603 (2007).

\bibitem{SC11}
R. Sharma and B.J. Cherayil,
Phys. Rev. E {\bf 83}, 041805 (2011).

\bibitem{LHS14}
F. Latinwo, K.-W Hsiao, and C.M. Schroeder, 
J. Chem. Phys. \textbf{141}, 174903 (2014).

\bibitem{VTC15}
M. Vucelja, K.S. Turitsyn, and M. Chertkov,
Phys. Rev. E {\bf 91}, 022123 (2015).

\bibitem{CG08}
R. Chetrite and K. Gaw\c{e}dzki,
Commun. Math. Phys. {\bf 282}, 469 (2008).

\bibitem{KP99}
P.E. Kloeden, and E. Platen,
\textit{Numerical Solutions of Stochastic Differential Equations}
(Springer, Berlin, Germany, 1999).

\end{thebibliography}
\end{document}